\documentclass[12pt,a4paper]{article}

\usepackage{amsfonts}
\usepackage{amsmath}

\def\Ro{{\mathbb R}}
\def\Co{{\mathbb C}}
\def\Io{{\mathbb I}}
\def\kk{{\bf k}}
\def\qq{{\bf q}}

\renewcommand{\Im}{\,{\rm Im}\,}

\begin{document}

\title{Photon fields in a fluctuating spacetime}

\author{  Jan Naudts$^\circ$, Maciej Kuna$^\dagger$,\\
          and Wojciech De Roeck$^\circ$\\
          \small
          $^\circ$Departement Natuurkunde, Universiteit Antwerpen UIA,\\
          \small
          Universiteitsplein 1, 2610 Antwerpen, Belgium\\
          \small
          $^\dagger$Wydzia\l\ Fizyki Technicznej i Matematyki Stosowanej,\\
	  \small
	  Politechnika Gda\'{n}ska,\\
	  \small
	  ul. Narutowicza 11/12, 80-952 Gda\'{n}sk, Poland\\
	  \small
	  E-mail: jan.naudts@ua.ac.be,
	  maciek@mifgate.mif.pg.gda.pl,\\
	  \small
	  wojciech.deroeck@ua.ac.be
}
\date{v5, October 2002}
\maketitle

\begin{abstract}

We present a model of interacting quantum fields, formulated in
a non-perturbative manner. One of the fields is treated
semi-classically, the other is the photon field. The
model has an interpretation of an electromagnetic field
in a fluctuating spacetime.

The model is equivalent with the quantization of 
electromagnetism proposed recently by Czachor.
Interesting features are that standard photon theory is recovered
as a limiting case, and that localized field operators
for the electromagnetic field exist as unbounded
operators in Hilbert space.

\end{abstract}

\section{Introduction}

\subsection{Two views on quantum spacetime}

At the heart of quantum mechanics are the dual concepts of 
states and observables. One way to describe quantum spacetime is 
by means of a non-abelian algebra of functions of position, or 
by introduction of non-commuting position-time observables. 
Proposals for the latter go back to the early days of quantum 
mechanics \cite{SHS47}. This kind of approach is also the basis 
for application of non-commutative geometry \cite{CA94}. The 
alternative, studied in the present paper, assumes that quantum 
spacetime has a state which is described by a vector $\Omega$ in 
some Hilbert space $\cal H$. This assumption implies that empty 
spacetime has some structure, made up by fluctuations imposed on 
top of Minkowski space. This structure is often referred to as 
the spacetime foam. The concept of a fluctuating spacetime goes 
back to the work of Wheeler \cite{WJA57}.

Let us assume that a shift of frame of the 
observer by a vector $q$ in Minkowski space is implemented by a 
unitary operator $U(q)$. The shift modifies the state vector $\Omega$ 
into the new vector $U(-q)\Omega$. One immediately expects
non-trivial autocorrelations of the form
\begin{equation}
w(q)=\langle\Omega|\,\hat U(-q)\Omega\rangle.
\label{autocor}
\end{equation}
However, since we cannot measure properties of spacetime without
putting particles in it, $w(q)$ is not directly observable.
In the standard concept of spacetime the vector $\Omega$
equals the vacuum vector $\Omega_0$. The latter is
invariant under translations. Hence, in that case $w(q)$
is constant equal to 1.
If on the other hand a spacetime structure exists on
the scale of Planck's length, or some other relevant
length scale $\mathit l$, then $w(q)$ is non-trivial.
It equals 1 for $q=0$ and vanishes for displacements,
large compared to $\mathit l$. In Fourier space,
these short-distance correlations lead to a cutoff
at high momenta.

\subsection{Discussion}

As is well known, the presence of a high-momentum
cutoff implies breaking of Poincar\'e invariance of the
field theory. In the present case
any Poincar\'e transformation $\Lambda$ maps the
state $\Omega$ of spacetime onto another state $\Lambda\Omega$.
This means that Poincar\'e covariance is maintained.
Poincar\'e invariance is restored at distances large compared to
$\mathit l$ by the assumption $w(q)\simeq 0$.
Experimental evidence for breaking of
Poincar\'e invariance has been analysed using
standard theory \cite {CG99}. If such a symmetry
breaking exists it should be extremely small,
because a very small symmetry violation
can produce dramatic observable effects.

The presence of a spacetime structure on
short length scales has effects similar to those
of noncommuting position-time operators.
Indeed, a high-momentum cutoff implies difficulties
for measurements at short length scales.
However, here, as in most quantum field theories,
it is not meaningful to define
position operators $Q_\mu$.
Instead, one can determine field strengths
at positions $q$ in Minkowsk space (see below).
Quantum expectations of these field strengths
are affected by spacetime fluctuations.
But it is difficult to conclude from inspection of these
measurable quantities that spacetime should be
noncommutative.

In early attempts to quantize gravity the metric
tensor $g_{\mu\nu}$ is treated as a potential,
analoguous to the vector potential of eletromagnetism.
The gravitons resulting from second quantization
are then treated in the same way as other field particles.
In particular, gravitons are expected to interact with all other
particles. In the present model the interplay between
spacetime fluctuations and field particles is more fundamental
and cannot be described in terms of interactions.
Field particles like photons or electrons cannot
exist without spacetime fluctuations. The structure
of the total system is far from the usual tensor product
description of independent subsystems, which are brought
into interaction via lagrangian or hamiltonian constraints.

Another line of reasoning predicts fluctuations in the speed
of light due to quantum uncertainties of matter fields.
This has been worked out in detail by Ford et al
\cite {FLH94,FS96}. Here, speed of light is constant equal to 1.
However, we still have to investigate the effect of an inhomogeneous state
of spacetime on the effective propagation of electromagnetic radiation.

\subsection {Czachor's noncanonical photon theory}

As noted by 
several authors, there is some inconsistency between a 
description of the electromagnetic field by means of state 
vectors in a bosonic Fock space, with a trivial vacuum state, and 
the regression to wavefunctions that are ground states of the 
harmonic oscillator. Czachor et al \cite {CM00,CS01,CM02} have proposed to solve 
this discord by modifying the link between harmonic oscillators 
and photons. As a first step, they replace all harmonic 
oscillators by a single oscillator with quantized frequency, the 
eigenvalues of which are the photon frequencies. In the next 
step, several copies of this quantized oscillator are introduced.
Photons become collective excitations of a spacetime filled 
with oscillators.

Features of Czachor's theory are that creation and annihilation
operators satisfy generalized commutation relations, and
that the vacuum state is not any longer unique.
The present paper starts from rather different assumptions
to arrive at the same theory, and hence,
the same physics.
In particular, we know from Czachor's work that
the standard theory of quantized free
electromagnetic fields is a limiting case of the present theory.
Another property of this theory is that field operators
$\hat A(q)$ for photons or electrons exist as genuine
operators in Fock space, and not as operator valued distributions,
as is the case in the conventional theory.
The latter is important for locality of
the electron-photon interaction. In the present formalism, the interaction term
\begin{equation}
\int_{\Ro^4}\hbox{ d}q\,\hat J^\mu(q) \hat A_\mu(q),
\end{equation}
with $J_\mu(q)$ the electron current,
exists as an operator in Hilbert space.

\subsection{Structure of the paper}

In the next section we formulate our model guided
by symmetry considerations. In section 3 a state of this model
is described by means of its correlation functions.
Some of its properties are discussed.
In section 4 the link is made with the work of Czachor.
The appendices contain a definition of covariance systems
and some technical proofs.

\section{A model of quantum spacetime}

\subsection{Symmetry considerations}

An important symmetry of quantum fields is additivity of the fields.
This symmetry is central to the Weyl approach. Let $H,+$
denote the additive group of single-particle wavefunctions.
Then Fock space is a projective representation of this group.
Indeed, the Weyl operators $\hat W(\psi)$ satisfy the Weyl
form of the canonical commutation relations
\begin{equation}
\hat W(\phi)\hat W(\psi)=\hat W(\phi+\psi)\zeta(\phi,\psi)
\label{Weylccr}
\end{equation}
with cocycle $\zeta$ given by
\begin{equation}
\zeta(\phi,\psi)=\exp\left(i\Im\langle\phi|\,\psi\rangle\right),
\end{equation}
and
\begin{equation}
\langle\phi|\,\psi\rangle=\int_{\Ro^3}\hbox{ d}\kk\,\frac{1}{2|\kk|}
\overline{\phi(\kk)}\psi(\kk).
\end{equation}
The field operator $\hat A(\psi)$ is now the generator of $\hat W(\psi)$
\begin{equation}
\hat W(\psi)=\exp\left(i\hat A(\psi)\right).
\label{fieldopdef}
\end{equation}
The creation and annihilation operators are expressed in terms
of the field operators in the usual way
\begin{equation}
\hat A_\pm(\psi)=\frac{1}{2}\hat A(\psi)\pm\frac{i}{2}\hat A(i\psi).
\label{creandef}
\end{equation}

In the theory, presented below, additivity of the field of spacetime
fluctuations is a broken symmetry, as a consequence of
the presence of additional particle fields. Hence, the Weyl approach
is not very well suited to describe fields of spacetime
fluctuations. This problem refrains us from formulating a theory
system based on the product of two additive groups, the group of
fields of spacetime fluctuations, and the group of additional
particle fields. Instead, we are forced to describe spacetime fluctuations
by means of an algebra of observables ${\cal A}$. Its choice
will be discussed later on. For the description of
the electromagnetic field we can continue to use the Weyl
operator approach as discussed above.

We do not include the 
Poincar\'e group in the initial formulation of the theory. But 
this implies that Poincar\'e covariance has to be checked later 
on a per state basis.

\subsection{Covariance systems}

It is obvious to assume that the fluctuations of spacetime
have to be second-quantized, and that they form a scalar boson field.
Alternatively, a vector or tensor field can be postulated.
But this would complicate the theory. For the moment we prefer
to keep the model as simple as possible.

We use the correlation function approach of \cite {NK01}.
It starts from a covariance system, also known as $C^*$-dynamical
system. States of this system are determined by correlation functions.
By means of a generalized G.N.S.-theorem they determine
a Hilbert space representation. See Appendix A.
The approach is rather analoguous to that of Wightman functions \cite {WAS56}.
It has been tested in a number of cases \cite {NK00,NJ02}.
In particular, it has been used to give a clean treatment
of the theory of free photons \cite {NK02}.

As discussed earlier, the spacetime fluctuations will
be described by an algebra $\cal A$ of observables,
the electromagnetic field by an additive group $G$ of classical
fields, or equivalently (see \cite {NK02}), of test functions.
This two components can be combined into a covariance
system by letting $G$ act on $\cal A$
in a trivial manner. Hence, the covariance system of
choice is of the form $({\cal A}, G, \Io)$.

An obvious choice of algebra $\cal A$ would be the algebra of 
bounded operators in the Fock space of a scalar boson. However, 
for technical reasons we are forced to take the abelian 
subalgebra of functions of momenta. A consequence 
is that in the present model the spacetime fluctuations are 
still described in a semi-classical way.

More precisely, the $C^*$-algebra $\cal A$ consists of
functions $f^{(n)}(\kk_1,\kk_2,\cdots\kk_n)$
which depend on the number of particles $n$, and, for a given $n$, on
wavevectors $\kk_m$, $m=1,2,\cdots n$ in $\Ro^3$.
Because the particles are indistinguishable, invariance
under permutation of the arguments $\kk_1,\kk_2,\cdots\kk_n$
is assumed. Each function $f$ of $\cal A$ is represented
as an operator $\hat f$ in the Fock space of the scalar boson
given by
\begin{eqnarray}
\hat f\chi^{(n)}(\kk_1,\kk_2,\cdots\kk_n)
&=&f^{(n)}(\kk_1,\kk_2,\kk_2,\cdots\kk_n)
\chi^{(n)}(\kk_1,\kk_2,\cdots\kk_n).
\label{fhatdef}
\end{eqnarray}
The norm $||f||$ of $f$ is then equal to the operator norm $||\hat f||$.

The symmetry group $G$ is the additive group of test functions
of the electromagnetic field.
It consists of complex continuous functions $\phi_\mu(\kk)$, with
compact support, depending on wavevectors in $\Ro^3$, and labeled with an
index running from 0 to 3. They satisfy the gauge condition
\begin{equation}
|\kk|\phi_0(\kk)=\sum_{\alpha=1}^3\kk_\alpha\phi_\alpha(\kk).
\label{psigauge}
\end{equation}
This group is locally compact in the discrete topology.

\section{States of the model system}

When describing interactions between particles it is
tradition to start from product states and then, to formulate
lagrangian or hamiltonian interactions between the particles.
However, for the states studied below the interaction is
so strong that it seems unlikely that a hamiltonian or
lagrangian description should exist.

\subsection{Product states}

The vacuum-to-vacuum correlation functions of
the free electromagnetic field define a state of the
covariance system $(\Co,G,\Io)$ by (see \cite {NK02})
\begin{eqnarray}
{\cal F}(\phi;\psi)&=&
\exp\left(-i\Im\langle \psi|\phi\rangle
\right)\
\exp\left(-\frac{1}{2}\,\langle \psi-\phi|
\psi-\phi\rangle\right)
\label{corphot}
\end{eqnarray}
with the (degenerate) scalar product given by
\begin{equation}
\langle \psi|\phi\rangle =-\int_{\Ro^3}\hbox{ d}\kk\,
\frac{1}{2|\kk|}\, \overline {\psi^\mu(\kk)}
\phi_\mu(\kk)\ge 0.
\label{bilin}
\end{equation}
On the other hand, if $\chi$ a state vector in the Fock space
of a scalar boson field then a state of the $C^*$-algebra $\cal A$ is defined
by ${\cal F}_\chi(f)=\langle\chi|\,\hat f\chi\rangle$.
In the special case that $\chi$ is of the form
\begin{equation}
\chi=\chi^{(0)}\oplus\chi^{(1)}
\oplus\frac{1}{\sqrt{2!}}\chi^{(2)}\otimes\chi^{(2)}
\oplus\cdots,
\label{classicstate}
\end{equation}
then the state on $\cal A$ is given by
\begin{eqnarray}
{\cal F}_\chi(f)
&=&f^{(0)}|\chi^{(0)}|^2\cr
& &
+\sum_{n=1}^\infty\frac{1}{n!}\left[\prod_{j=1}^n\int_{\Ro^3}\hbox{ d}\kk_j
\frac{1}{2|\kk_j|}\,|\chi^{(n)}(\kk_j)|^2\right]
 f^{(n)}(\kk_1,\kk_2,\cdots,\kk_n).\cr
& &
\label{corfluct}
\end{eqnarray}

The product of (\ref{corphot}) and (\ref{corfluct}) defines
a state of the covariance system $({\cal A},G,\Io)$.
However, we are not interested in this product state.
In what follows another way of combining (\ref{corphot}) and (\ref{corfluct}) 
is investigated.

\subsection{Interacting states}

An alternative way of combining (\ref{corphot}) and (\ref{corfluct})
is obtained by first bringing the integrations in the
exponential of (\ref{corphot}) in front of the whole expression
\begin{eqnarray}
& &\exp\left(-i\Im\langle \psi|\phi\rangle
-\frac{1}{2}\,\langle \psi-\phi|
\psi-\phi\rangle\right)\cr
&\Rightarrow&
\int_{\Ro^3}\hbox{ d}\kk
\exp\left(i\Im\overline{ \psi^\mu(\kk)}\phi_\mu(\kk)
+\frac{1}{2}\,\overline{( \psi^\mu(\kk)-\phi^\mu(\kk))}
(\psi_\mu(\kk)-\phi_\mu(\kk))
\right)\cr
& &
\end{eqnarray}
and next modifying all integrations in (\ref{corfluct})
so as to include the exponential integrand of the above expression.
As a final modification we introduce real constants $c_n$
which multiply the test functions $\phi$ in $G$, and make their
amplitude dependend on the number of spacetime fluctuations.
All together, the result is
\begin{eqnarray}
& &{\cal F}_\chi(f;\phi;\psi)
=f^{(0)}|\chi^{(0)}|^2\cr
& &
+\sum_{n=1}^\infty\frac{1}{n!}\bigg[\prod_{j=1}^n\int_{\Ro^3}\hbox{ d}\kk_j
\frac{1}{2|\kk_j|}\,|\chi^{(n)}(\kk_j)|^2
\exp\left(ic_n^2\Im\overline{ \psi^\mu(\kk_j)}\phi_\mu(\kk_j)
\right)\cr
& &\times\exp\left(
\frac{c_n^2}{2}\,\overline{( \psi^\mu(\kk_j)-\phi^\mu(\kk_j))}
(\psi_\mu(\kk_j)-\phi_\mu(\kk_j))
\right)
\bigg]\cr
& &\times f^{(n)}(\kk_1,\kk_2,\cdots,\kk_n).\cr
& &
\label{corfunext}
\end{eqnarray}

Covariance of the correlation functions can be checked easily. 
See Appendix B. The multiplier $\zeta$ appearing in the Weyl 
commutation relations (\ref{Weylccr}) for the electromagnetic 
field is modified into an operator $\hat\zeta$, which represents 
the function
\begin{equation}
\zeta(\phi;\psi)^{(n)}(\kk_1,\cdots,\kk_n)
=\exp\left(ic_n^2\sum_{j=1}^n\Im\overline{\psi^\mu(\kk_j)}\phi_\mu(\kk_j)\right).
\end{equation}

\subsection{Hilbert space representation}

The generalized GNS-theorem of \cite {NK01} implies the existence of
a representation of the algebra $\cal A$ as bounded operators of
a Hilbert space $\cal H$. The operator corresponding with the
function $f^{(n)}(\kk_1,\kk_2,\cdots,\kk_n)$ is denoted $\hat f$,
as before. Note however that the Hilbert space has changed so that
$\hat f$ is not any longer defined by (\ref{fhatdef}).
There exist also unitary operators $\hat W(\phi)$ in $\cal H$, one for each element
of the group $G$, with the property that
\begin{equation}
\hat W(\phi)\hat W(\psi)=
\hat W(\phi+\psi)
\hat\zeta(\phi;\psi)
\label{Weylccr2}
\end{equation}
Finally there exists in $\cal H$ a vector
$\Omega$, such that
\begin{eqnarray}
{\cal F}_\chi(f;\phi;\psi)
&=&\langle\Omega|\,\hat W(\psi)\hat f\hat W(\phi)^*\Omega\rangle
\label{repr}
\end{eqnarray}
holds for all choices of the fields $\phi$ and $\psi$.

Note that the operators $\hat f$ commute
with all operators $\hat W(\phi)$. This is a consequence of the fact that
the action of the group $G$ on the algebra $\cal A$ is trivial.
In other words, the algebra of operators
$\hat f$ belongs to the center of the representation.
The representation is reducible, which can be understood because
the present model treats the spacetime fluctuations in a semi-classical
manner.

From (\ref{Weylccr2}) follows that
\begin{equation}
\hat W(\phi)\hat W(\psi)=
\hat W(\psi)\hat W(\phi)\hat\zeta(\psi,\phi)^*\hat\zeta(\phi,\psi).
\end{equation}
Combining this expression with the definition of the field operators (\ref{fieldopdef})
one obtains the commutation relations
\begin{equation}
\left[\hat A(\psi),\hat A(\phi)\right]_-
=i\hat s(\psi,\phi),
\label{nccr}
\end{equation}
with
\begin{equation}
\hat\zeta(\phi,\psi)=\exp((i/2)\hat s(\psi,\phi)),
\end{equation}
where $\hat s(\psi,\phi)$ is the operator corresponding with the function
\begin{equation}
s^{(n)}(\psi,\phi)(\kk_1,\cdots,\kk_n)
=2c_n^2\sum_{j=1}^n\Im\overline {\psi^\mu(\kk_j)}\phi_\mu(\kk_j).
\end{equation}
The commutation relations (\ref{nccr}) differ from the traditional
result
\begin{equation}
\left[\hat A(\psi),\hat A(\phi)\right]_-
=-2i\Im\langle\psi|\,\phi\rangle
\end{equation}
because the r.h.s.~of the commutator is not a multiple of
the identity operator, but is an operator in the
center of the representation. This kind of commutation relations
has been studied before, in the context of
generalized free fields --- see e.g.~section 12.5 of \cite {BLT75}.

The operator $\hat A(\phi)$ is a real linear function of its
argument $\phi$ --- see Appendix C. It is also shown there that
\begin{equation}
\hat A(\phi)\Omega=i\hat A(i\phi)\Omega
\label{idep}
\end{equation}
holds for all $\phi$.
An immediate consequence is that the annihilation operator $\hat A_-(\phi)$,
defined by (\ref{creandef}), annihilates the state vector $\Omega$. Hence the latter acts
as a vacuum for the photon field. But notice that it is not a vacuum for the field
of spacetime fluctuations.

\subsection{Poincar\'e covariance}

A shift by a vector $q$ of Minkowski space leaves functions of momentum
invariant and maps the state vector $\Omega$ into the vector $\hat u_{-q}\Omega$,
where $\hat u_q$ is the representation of the function
\begin{eqnarray}
u_q^{(n)}(\kk_1,\cdots,\kk_n)=e^{-iq_0\sum_{j=1}^n|\kk_j|}e^{i\qq\cdot\sum_{j=1}^n\kk_j}.
\end{eqnarray}
However, a shift in Minkowski space affects also the
electromagnetic field. Test functions $\phi$ in $G$
transform according to
\begin{equation}
\tau_q\phi_\mu(\kk)=
\phi_\mu(\kk)\exp(iq_0|\kk|)\exp(-i\qq\cdot\kk).
\end{equation}
Therefore, the unitary operator $\hat U(q)$, implementing a shift by $q$ in Minkowski space,
is defined by
\begin{equation}
\hat U(q)\hat f\hat W(\phi)\Omega=\hat f\hat u_{q}\hat W(\tau_q\phi)\Omega.
\end{equation}
It is straightforward to verify that these operators $\hat U(q)$ form
a unitary representation of $\Ro^4,+$.

Calculate now
\begin{eqnarray}
w(q)&=&\langle\Omega|\,\hat U(-q)\Omega\rangle\cr
&=&{\cal F}_\chi(u_{-q};0;0)\cr
&=&|\chi^{(0)}|^2\cr
& &
+\sum_{n=1}^\infty\frac{1}{n!}\bigg[\int_{\Ro^3}\hbox{ d}\kk
\frac{1}{2|\kk|}\,|\chi^{(n)}(\kk)|^2\exp(iq_0|\kk|)\exp(-i\qq\cdot\kk)
\bigg]^n.\cr
& &
\end{eqnarray}
From this expression it is clear that for large values of $q$ the
correlation function $w(q)$ tends to $|\chi^{(0)}|^2$,
while for $q=0$ one has $w(0)=1$.

The generators of the group of unitary shift operators are the
momentum operators $\hat K_\mu$, defined by
\begin{equation}
\hat U(q)=\exp(-iq^\mu\hat K_\mu).
\label{defgenshift}
\end{equation}
In particular, $\hat K_0$ is the energy operator.
A short calculation gives
\begin{eqnarray}
\langle\Omega|\,\hat W(\psi)\hat f \hat U(q)\hat W(\phi)^*\Omega\rangle
&=&{\cal F}_\chi(fu_{q};\tau_q\phi;\psi).
\end{eqnarray}
Linearization in $q_0$ gives
\begin{eqnarray}
\langle\Omega|\,\hat W(\psi)\hat f \hat K_0\hat W(\phi)^*\Omega\rangle
&=&\sum_{n=1}^\infty
\left[\left[
\sum_{l=1}^n|\kk_l|
\left(1-c_n^2\overline{\phi^\mu(\kk_l)}\psi_\mu(\kk_l)\right)
\right]\right]_n
\end{eqnarray}
with
\begin{eqnarray}
\left[\left[
X\right]\right]_n
&=&
\frac{1}{n!}\bigg[\prod_{j=1}^n\int_{\Ro^3}\hbox{ d}\kk_j
\frac{1}{2|\kk_j|}\,|\chi^{(n)}(\kk_j)|^2\cr
& &\times
\exp\left(ic_n^2\Im\overline{ \psi^\mu(\kk_j)}\phi_\mu(\kk_j)
\right)\cr
& &\times\exp\left(
\frac{c_n^2}{2}\,\overline{( \psi^\mu(\kk_j)-\phi^\mu(\kk_j))}
(\psi_\mu(\kk_j)-\phi_\mu(\kk_j))
\right)
\bigg]\cr
& &\times f^{(n)}(\kk_1,\kk_2,\cdots,\kk_n)
\,X.\cr
& &
\end{eqnarray}
This result can be used to show that $\hat K_0$ is a positive operator.
It has two contributions, the energy of the spacetime fluctuations,
and that of the electromagnetic field.
A similar calculation shows that also the momenta $\hat K_\alpha,\alpha=1,2,3$
have two contributions,
one of the spacetime fluctuations, the other of the electromagnetic field.
Further calculations show that the squared mass operator satisfies
$\hat K^\mu \hat K_\mu\ge 0$. We omit these calculations here.

Under a Lorentz transformation $\Lambda$ a 3-dimensional wave vector $\kk$
transforms into the vector $\kk'$ with components
\begin{equation}
\kk'_\alpha=(\Lambda^{-1})_\alpha^{\,0}|\kk|
+\sum_{\beta=1}^3(\Lambda^{-1})_\alpha^{\,\beta}\kk_\beta.
\end{equation}
The state with state vector $\chi$ and correlation functions (\ref{corfunext}) transforms
into a state described by the statevector $\chi'$ with components
\begin{equation}
{\chi'}^{(n)}(\kk)=\chi^{(n)}(\kk').
\end{equation}
The Lorentz transformations are not necessarily unitarily implemented
in the present representation.

\subsection{Infinite number of spacetime fluctuations} 

Take now $c_n=1/\sqrt n$ and consider a state vector $\chi$ of the form (\ref{classicstate}),
with $\chi^{(n)}(\kk)=\delta_{n,N}(N!)^{1/2N} Z(\kk)^{1/2}$, and with $Z(\kk)$
a normalized density function
\begin{equation}
\int_{\Ro^3}\hbox{ d}\kk\,\frac{1}{2|\kk|}\,Z(\kk)=1.
\end{equation}
Then one calculates
\begin{eqnarray}
& &\lim_{N\rightarrow\infty}
{\cal F}_\chi(1;\phi;\psi)\cr
&=&\lim_{N\rightarrow\infty}
\bigg[\int_{\Ro^3}\hbox{ d}\kk\,\frac{1}{2|\kk|}\,Z(\kk)
\exp\left((i/N)\Im\overline{ \psi^\mu(\kk)}\phi_\mu(\kk)
\right)\cr
& &\times\exp\left(
\frac{1}{2N}\,\overline{( \psi^\mu(\kk)-\phi^\mu(\kk))}
(\psi_\mu(\kk)-\phi_\mu(\kk))
\right)
\bigg]^N\cr
&=&\exp\left(i\Im
\int_{\Ro^3}\hbox{ d}\kk\,\frac{1}{2|\kk|}\,Z(\kk)
\overline{ \psi^\mu(\kk)}\phi_\mu(\kk)\right)\cr
& &\times
\exp\left(
\frac{1}{2}\int_{\Ro^3}\hbox{ d}\kk\,\frac{1}{2|\kk|}\,Z(\kk)
\overline{( \psi^\mu(\kk)-\phi^\mu(\kk))}
(\psi_\mu(\kk)-\phi_\mu(\kk))
\right).\cr
& &
\end{eqnarray}
This expression coincides with result (\ref{corphot}) of the standard
photon theory, except for the appearance of the function $Z(\kk)$,
which acts as a cutoff for large $\kk$-values.

\section{Explicit construction \`a la Czachor}

\subsection{A single spacetime fluctuation}

In the approach of \cite {CM00,CS01,CM02} photons in presence of
a single spacetime fluctuation are described by a pair of
harmonic oscillators, the frequency of which 
has been quantized and has become an operator. The excitations
of the oscillators are the photons. Two oscillators
are needed because of the two polarizations of the photon.
The frequency of the photon can equal any of the
eigenvalues of the frequency operator.

Let $\hat a$ and $\hat a^\dagger$ be the annihilation
and creation operators of the standard harmonic oscillator. They satisfy
the commutation relations
\begin{equation}
\left[\hat a,\hat a^\dagger\right]=1.
\end{equation}
The Hilbert space of wavefunctions consists of functions of the form
$\chi(\kk,m,n)$, where $\kk$ is a wavevector and $m$ and $n$ are
quantum numbers of the two harmonic oscillators. Normalization
is such that
\begin{equation}
\sum_{m,n=0}^{+\infty}\int_{\Ro^3}\hbox{ d}\kk\,\frac{1}{2|\kk|}
\,|\chi(\kk,m,n)|^2=1.
\end{equation}
Any wavefunction of the form
\begin{equation}
\chi(\kk,m,n)=\chi(\kk)\delta_{m,0}\delta_{n,0}
\end{equation}
is a vacuum vector for the photons and will be denoted
\begin{equation}
|\chi\rangle\otimes|0\rangle\otimes|0\rangle.
\end{equation}
For any bispinor
\begin{equation}
\phi =\left(\begin{array}{c}
\phi _1 \\
\phi _2 \\
\end{array}\right),
\end{equation}
consisting of two test functions $\phi_1(\kk)$ and $\phi_2(\kk)$,
one defines an annihilation operator $\hat A_-(\phi )$ by
\begin{equation}
\hat A_-(\phi )=\hat \phi _1\otimes \hat a\otimes\Io
+\hat \phi _2\otimes \Io\otimes \hat a
\label{amindef}
\end{equation}
(the operators $\hat \phi _1,\hat \phi _2$ are defined as multiplication
operators).
The creation operator $\hat A_+(\phi )$ is the conjugate
of the annihilation operator,
as usual. These creation and annihilation operators satisfy
the commutation relations
\begin{equation}
\left[\hat A_-(\psi ),\hat A_+(\phi )\right]
=\left(\hat \psi _1\hat \phi _1^*+\hat \psi _2\hat \phi ^*_2\right)\otimes \Io\otimes \Io.
\label{cznccr1}
\end{equation}
The field operators $\hat A(\phi )=\hat A_-(\phi )+\hat A_+(\phi )$
satisfy
\begin{eqnarray}
\left[\hat A(\psi ),\hat A(\phi )\right]
&=&\left(\hat \psi _1\hat \phi _1^*+\hat \psi _2\hat \phi ^*_2
-\hat \phi _1\hat \psi _1^*-\hat \phi _2\hat \psi ^*_2
\right)\otimes \Io\otimes \Io.
\label{cznccr2}
\end{eqnarray}
Comparison of (\ref{cznccr2})
with (\ref{nccr}) is not immediately possible because here two-component
spinors are used instead of four-component test functions.

\subsection{Localized field operators}

Following \cite {CM02}, momentum operators $\hat K_\mu$ are defined by
\begin{eqnarray}
\hat K_0\phi(\kk,m,n)
&=&|\kk|\left(\frac{1}{2}+m\right)\left(\frac{1}{2}+n\right)\phi(\kk,m,n)\cr
\hat K_\alpha\phi(\kk,m,n)
&=&\kk_\alpha\left(\frac{1}{2}+m\right)\left(\frac{1}{2}+n\right)\phi(\kk,m,n).
\end{eqnarray}
They define unitary shift operators  $\hat U(q)$ by (\ref{defgenshift}).

It is now immedately clear from (\ref{amindef}) that the
localized annihilation operators $\hat A_-^{(\rm l)}(q)$ and
$\hat A_-^{(\rm r)}(q)$, one for each of the two polarizations of the photon,
are given by
\begin{eqnarray}
\hat A_-^{(\rm l)}(q)&=&e^{-iq^\mu\hat K_\mu}\otimes\hat a\otimes\Io,\cr
\hat A_-^{(\rm r)}(q)&=&e^{-iq^\mu\hat K_\mu}\otimes\Io\otimes\hat a.
\end{eqnarray}

\subsection{Correlation functions}

The field operators $\hat A(\phi )$
are used to define unitary operators $\hat W(\phi )$ by
\begin{eqnarray}
\hat W(\phi )
&=&\exp\left(i\hat A(\phi )\right).
\end{eqnarray}
The Weyl operators $\hat W(\phi )$ satisfy the product rule
\begin{equation}
\hat W(\phi )\hat W(\psi )=\left(e^{(i/2)\hat s(\phi ,\psi )}\otimes\Io\otimes\Io\right)
\hat W(\phi +\psi )
\end{equation}
with
\begin{equation}
s(\phi ,\psi )=-2\Im\left(\phi _1\overline{\psi _1}+\phi _2\overline{\psi _2}\right).
\end{equation}
The latter result is obtained by means of the Campbell-Baker-Hausdorff
formula
\begin{equation}
\exp\left(i(A+B)\right)=\exp\left(iA\right)\exp\left(iB\right)
\exp\left((1/2)[A,B]\right),
\end{equation}
which is valid when $B$ commutes with $[A,B]$.

The action of $\hat W(\phi )$ on a vacuum vector is given by
\begin{equation}
\hat W(\phi )|\chi\rangle\otimes|0\rangle\otimes|0\rangle(\kk)
=\chi(\kk)
e^{i(\phi _1(\kk)\hat a+\overline{\phi _1(\kk)}\hat a^\dagger)}|0\rangle
\otimes
e^{i(\phi _2(\kk)\hat a+\overline{\phi _2(\kk)}\hat a^\dagger)}|0\rangle.
\end{equation}
Using the relation
\begin{equation}
e^{\lambda\hat a+\mu\hat a^\dagger}|0\rangle
=e^{(1/2)\lambda\mu}\sum_n\frac{1}{\sqrt{n!}}\mu^n|n\rangle,
\end{equation}
the above expression can be rewritten as
\begin{eqnarray}
\hat W(\phi )|\chi\rangle\otimes|0\rangle\otimes|0\rangle(\kk)
&=&\chi(\kk)
\exp\left(-\frac{1}{2}|\phi _1(\kk)|^2-\frac{1}{2}|\phi _2(\kk)|^2\right)\cr
& &\times
\sum_{m,n}\frac{i^{m+n}}{\sqrt{m!n!}}\overline{\phi _1(\kk)^m\phi _2(\kk)^n}
|m\rangle\otimes|n\rangle.
\end{eqnarray}

The correlation functions are now calculated as follows.
\begin{eqnarray}
{\cal F}_\chi(f;\phi;\psi)
&=&\langle\chi|\otimes\langle 0|\otimes\langle 0|
\,\hat W(\psi)\hat f\hat W(\phi)^*
\chi\rangle\otimes |0\rangle\otimes |0\rangle\cr
&=&\int_{\Ro^3}\hbox{ d}\kk\frac{1}{2|\kk|}f(\kk)|\chi(\kk)|^2\cr
&\times&\prod_{s=1,2}
\exp(-(1/2)|\phi_s(\kk)|^2-(1/2)|\psi_s(\kk)|^2-\psi_s(\kk)\overline{\phi_s(\kk)}).\cr
& &
\end{eqnarray}
Again, this formula does not compare directly with (\ref{corfunext})
because here two-component spinors are used instead of four-component classical wavefunctions.

\subsection{Multiple fluctuations}

The transition from the one-oscillator to the multi-oscillator formalism
is straightforward. Starting from the Hilbert space $\cal H$ of wave functions
$\chi(\kk,m,n)$ a Fock space $\underline{\cal H}$ is constructed in the
usual way. The $n$-th component of $\underline{\cal H}$ is the space
$\otimes^n_{\rm s}{\cal H}$ of these elements of ${\cal H}^{\otimes n}$
which are symmetric under permutations. The field operator
$\underline{\hat A}(\phi)$, acting on $\underline{\cal H}$,
is defined by
\begin{eqnarray}
\underline{\hat A}(\phi)=0\oplus\hat A(\phi)
\oplus\frac{1}{\sqrt 2}\left(\hat A(\phi)\otimes\Io+\Io\otimes \hat A(\phi)\right)
+\cdots.
\end{eqnarray}
Consider now a vacuum vector in $\cal H$ of the form
\begin{equation}
\Omega=|\chi\rangle\otimes|0\rangle\otimes|0\rangle.
\end{equation}
Fix a discrete probability distribution $p$ i.e., positive numbers $p_n$
satisfying $\sum_np_n=1$. Then a vacuum vector $\underline\Omega$ of $\underline{\cal H}$
is defined by
\begin{eqnarray}
\underline\Omega
&=&\sqrt{p_0}\oplus \sqrt{p_1}\Omega\oplus \sqrt{p_2}\Omega\otimes\Omega\oplus\cdots.
\end{eqnarray}
Correlation functions can now be calculated as follows.
\begin{eqnarray}& &
{\cal F}_\chi(f;\phi;\psi)\cr
&=&\langle\underline\Omega|
\,\underline{\hat W}(\psi)\hat f\underline{\hat W}(\phi)^*
\underline\Omega\rangle\cr
&=&p_0f^{(0)}+\sum_{n=1}^\infty p_n
\bigg[\prod_{j=1}^n
\int_{\Ro^3}\hbox{ d}\kk_j\frac{1}{2|\kk_j|}|\chi(\kk_j)|^2\cr
&\times&\prod_{s=1,2}
\exp\left(-\frac{1}{2n}(|\phi_s(\kk_j)|^2+|\psi_s(\kk_j)|^2
+2\psi_s(\kk_j)\overline{\phi_s(\kk_j)})\right)
\bigg]\cr
& &\times f^{(n)}(\kk_1,\cdots,\kk_n).
\end{eqnarray}
Comparison with (\ref{corfunext}) shows that here the constants $c_n$ equal $1/\sqrt n$.
The state vector $\chi$ in the Fock space of the scalar boson is of
the special form (\ref{classicstate}),
with $\chi^{(n)}(\kk)=n!p_n\chi(\kk)$. In \cite {CM02}, the special choice
\begin{equation}
p_n=e^{-\alpha}\frac{\alpha^n}{n!}
\end{equation}
is discussed. It is argued that it restores Poisson statistics of the photons
in the limit of an infinite number of spacetime fluctuations.

\section*{Appendix A}

The {\sl covariance system} $({\cal A},G,\sigma)$ consists of a $C^*$-algebra
$\cal A$, a locally compact group $G$, and an action $\sigma$ of $G$
as automorphisms of $\cal A$. It is assumed that the map
$g\in G\rightarrow \sigma_g f$ is continuous for each $f$ in $\cal A$.
A {\sl state} of the covariance system is determined by correlation
functions ${\cal F}(f;g;g')$ with $f$ in $\cal A$ and $g,g'$ in $G$.
They should satisfy the following conditions (for simplicity
we assume that $\cal A$ is a $C^*$-algebra with unit)
\begin{description}
\item {}(positivity) For all $n>0$ and for all possible choices of complex $\lambda_j$,
$g_j\in G$, and $f_j\in{\cal A}$, is
\begin{equation}
\sum_{j,k=1}^n\lambda_j\overline{\lambda_k}{\cal F}(f_k^*f_j;g_j;g_k)\ge 0;
\end{equation}
\item {}(normalization) ${\cal F}(1;e;e)=1$, where $1$ is the unit of $\cal A$,
and $e$ is the neutral element of $G$;
\item {}(covariance) there exists a right multiplier $\zeta$ with values in $\cal A$
such that
\begin{equation}
{\cal F}(\sigma_{g''}f;g;g')={\cal F}(\zeta(g',g'')f\zeta(g,g'')^*;gg'';g'g'');
\label{covdef}
\end{equation}
\item {}(continuity) for each $f$ in $\cal A$ the map $g,g'\rightarrow {\cal F}(f;g;g')$ is
continuous in a neighborhood of the neutral element $e,e$.
\end{description}

The first two conditions imply that the map $f\in{\cal A}\rightarrow {\cal F}(f;e;e)$
defines a state of the $C^*$-algebra $\cal A$.

\section*{Appendix B}

Here we show that the correlation functions(\ref{corfunext})
define a state of the covariance system $({\cal A},G,\Io)$.

\paragraph{Positivity}
One calculates
\begin{eqnarray}
& &\sum_{m,l}\lambda_m\overline{\lambda_l}{\cal F}_\chi(f^*_lf_m;\phi_m;\phi_l)\cr
&=&|\sum_m\lambda_mf_m^{(0)}|^2\,|\omega^{(0)}|^2\cr
& &+
\sum_{n=1}^\infty\frac{1}{n!}\left[\prod_{j=1}^n\int_{\Ro^3}\hbox{ d}\kk_j
\frac{1}{2|\kk_j|}\,|\chi^{(n)}(\kk_j)|^2\right]\cr
& &\times
\sum_{m,l}\mu_m(\kk_1,\cdots\kk_n)\overline{\mu_l}(\kk_1,\cdots,\kk_n)
\exp\left(-c_n^2\sum_{j=1}^n\overline{\phi_m^\mu(\kk_j)}\phi_{l,\mu}(\kk_j)\right)\cr
& &
\label{appa1}
\end{eqnarray}
with
\begin{equation}
\mu_m^{(n)}(\kk_1,\cdots,\kk_n)
=\lambda_m
f_m^{(n)}(\kk_1,\cdots,\kk_n)
\exp\left(\frac{c_n^2}{2}\sum_{m=1}^n\overline{\phi_m^\mu(\kk_m)}\phi_{m,\mu}(\kk_m)\right).
\end{equation}
Note that the matrix $M$ with elements
\begin{equation}
M_{j,l}=-c_n^2\sum_{j=1}^n\overline{\phi_m^\mu(\kk_j)}\phi_{l,\mu}(\kk_j)
\end{equation}
is positive definite because the test functions $\phi_m(\kk)$ satisfy
(\ref{psigauge}). Hence, by Schur's lemma, also the matrix with
elements $\exp(M_{j,l})$ is positive  One concludes
that (\ref{appa1}) is positive.

\paragraph{Normalization}
is trivially satisfied provided that the vector $\chi$ in Fock space is
properly normalized.

\paragraph{Covariance}
One calculates
\begin{eqnarray}
& &{\cal F}(f;\phi+\xi;\psi+\xi)
=f^{(0)}|\chi^{(0)}|^2
+\sum_{n=1}^\infty\frac{1}{n!}\bigg[\prod_{j=1}^n\int_{\Ro^3}\hbox{ d}\kk_j
\frac{1}{2|\kk_j|}\,|\chi^{(n)}(\kk_j)|^2\cr
& &\times
\exp\left(ic_n^2\Im\overline{( \psi^\mu(\kk_j)+\xi^\mu(\kk_j))}
(\phi_\mu(\kk_j)+\xi_\mu(\kk_j))
\right)\cr
& &\times\exp\left(
\frac{c_n^2}{2}\,\overline{( \psi^\mu(\kk_j)-\phi^\mu(\kk_j))}
(\psi_\mu(\kk_j)-\phi_\mu(\kk_j))
\right)
\bigg]\cr
& &\times f^{(n)}(\kk_1,\kk_2,\cdots,\kk_n)\cr
&=&{\cal F}(\overline{\zeta(\psi,\xi)}f\zeta(\phi,\xi);\phi;\psi).
\end{eqnarray}
This proves covariance of the correlation functions.

\paragraph{Continuity}
The continuity requirement is empty because of the discrete topology
of the group $G$.

One concludes that the correlation functions (\ref{corfunext})
define a state of the covariance system $({\cal A},G,\Io)$.

\section*{Appendix C}

Here we show that the field operator $\hat A(\phi)$ is a real linear function
of its argument $\phi$. We also show that (\ref{idep}) holds.

One calculates
\begin{eqnarray}
& &\langle\Omega|\,\hat W(\psi)\hat W(\xi)\hat f\hat W(\phi)^*\Omega\rangle\cr
&=&\langle\Omega|\,\hat W(\psi+\xi)\hat\zeta(\psi,\xi)\hat f\hat W(\phi)^*\Omega\rangle\cr
&=&{\cal F}_\chi(\zeta(\psi,\xi)f;\phi;\psi+\xi).
\label{appb1}
\end{eqnarray}

First note that the map
\begin{equation}
\lambda\in\Ro\rightarrow
\langle\Omega|\,\hat W(\psi)\hat W(\lambda\xi)\hat f\hat W(\phi)^*\Omega\rangle
\end{equation}
is continuous. This suffices, together with
\begin{equation}
\hat W(\lambda\xi)\hat W(\mu\xi)=\hat W((\lambda+\mu)\xi),
\end{equation}
to apply Stone's theorem and to conclude the existence of self-adjoint operators
$\hat A(\xi)$ satisfying $\hat A(\lambda\xi)=\lambda\hat A(\xi)$
for all real $\lambda$.

Next, linearize (\ref{appb1}) in $\xi$, using (\ref{corfunext}). One obtains
\begin{eqnarray}
& &i\langle\Omega|\,\hat W(\psi)\hat A(\xi)\hat f\hat W(\phi)^*\Omega\rangle
=f^{(0)}|\chi^{(0)}|^2\cr
& &
+\sum_{n=1}^\infty\frac{c_n^2}{n!}\bigg[\prod_{j=1}^n\int_{\Ro^3}\hbox{ d}\kk_j
\frac{1}{2|\kk_j|}\,|\chi^{(n)}(\kk_j)|^2
\exp\left(ic_n^2\Im\overline{ \psi^\mu(\kk_j)}\phi_\mu(\kk_j)
\right)\cr
& &\times\exp\left(
\frac{c_n^2}{2}\,\overline{( \psi^\mu(\kk_j)-\phi^\mu(\kk_j))}
(\psi_\mu(\kk_j)-\phi_\mu(\kk_j))
\right)\bigg]\cr
& &\times\sum_{l=1}^n\left[\overline{\xi^\mu(\kk_l)}\psi_\mu(\kk_l)
-\overline{\phi^\mu(\kk_l)}\xi_\mu(\kk_l)\right]
\cr
& &\times f^{(n)}(\kk_1,\kk_2,\cdots,\kk_n).\cr
& &
\label{appb2}
\end{eqnarray}
This result shows that $\hat A(\xi)$ is a real linear function of $\xi$.

Finally, let us show (\ref{idep}).
Take $\psi=0$ in (\ref{appb2}). This gives
\begin{eqnarray}
& &i\langle\hat A(\xi)\Omega|\,\hat f\hat W(\phi)^*\Omega\rangle
=f^{(0)}|\chi^{(0)}|^2\cr
& &
-\sum_{n=1}^\infty\frac{c_n^2}{n!}\bigg[\prod_{j=1}^n\int_{\Ro^3}\hbox{ d}\kk_j
\frac{1}{2|\kk_j|}\,|\chi^{(n)}(\kk_j)|^2
\exp\left(
\frac{c_n^2}{2}\,\overline{\phi^\mu(\kk_j)}
\phi_\mu(\kk_j)
\right)\bigg]
\cr
& &\times 
\sum_{l=1}^n
\overline{\phi^\mu(\kk_l)}\xi_\mu(\kk_l)
f^{(n)}(\kk_1,\kk_2,\cdots,\kk_n).\cr
& &
\label{appb3}
\end{eqnarray}
This implies (\ref{idep})


\end{document}